\documentclass{jnmp}

\usepackage{amsmath}

\JNMPnumberwithin{equation}{section}

\theoremstyle{definition}

\begin{document}

%
\renewcommand{\evenhead}{P G Est\'evez and J Prada}
\renewcommand{\oddhead}{Singular manifold method for an equation in $2+1$ dimensions}

%
\thispagestyle{empty}

\FirstPageHead{*}{*}{20**}{\pageref{firstpage}--\pageref{lastpage}}{Article}

\copyrightnote{200*}{P G Est\'evez and J Prada}

\Name{Singular manifold method for an equation in $2+1$ dimensions}

\label{firstpage}

\Author{P. G. ESTEVEZ~$^\dag$ and J.  PRADA~$^\ddag$}

\Address{$^\dag$ Area de Fisica Te\'orica, Facultad de Ciencias, Universidad de
Salamanca, SPAIN \\
~~E-mail: pilar@usal.es\\[10pt]
$^\ddag$ Departamento de Matem\'aticas, Facultad de Ciencias, Universidad de
Salamanca, SPAIN \\
~~E-mail: prada@usal.es}

\Date{Received Month *, 200*; Revised Month *, 200*;
Accepted Month *, 200*}

\begin{abstract}
\noindent The Singular Manifold Method is presented as an
excellent tool to study a $2+1$ dimensional equation in despite of
the fact that the same method presents several problems when
applied to $1+1$ reductions of the same equation. Nevertheless
these problems are solved when the number of dimensions of the
equation is increased.
\end{abstract}

%
\section{Introduction}
There are many different approaches to the study of nonlinear partial differential equations. Equations in
$1+1$ dimensions  are considered as the easiest part of the field. The different methods
 start usually with equations in $1+1$ dimensions and then, when the method has succeeded,
 the generalization to $2+1$ dimensions can be considered \cite{konopel92}.

 Among the different methods to study a given PDE, the Singular Manifold Method (SMM)
 \cite{Weiss}
 based in the Painlev\'e property  \cite{WTC} has been proved to be very effective. As it is well known, the
 Painlev\'e test  is an algorithmic procedure that allows us to determine if the solutions of a PDE are
 singlevalued in the initial conditions. Essentially, for a PDE in $z_1...z_n$ variables, the Painlev\'e test
 requires that all the solutions of the PDE could be locally written as:
 \begin{equation}
 u(z_1,....z_n)=\sum_{j=0}^{\infty}u_j(z_1,....z_n)\left[\phi(z_1,....z_n)\right]^{j-a},\end{equation}
 where $a$ is an integer positive number and $\phi(z_1,....z_n)$ a totally arbitrary differentiable function.
 Furthermore, once the Painlev\'e test has been checked for a given PDE, the SMM allows us to derive B\"acklund
 transformations, Lax pair, Darboux transformations and tau-functions for the PDE.
 Nevertheless we must remember that there are some problems related with Painlev\'e property,  Painlev\'e test and SMM. We list some of them:
 \begin{itemize}
 \item One of the main criticisms to the Painlev\'e property is the fact that it is noninvariant under
 changes of dependent and/or independent variables. Many times it is not easy to identify the change of
 variables that allow us to write a PDE in a form such that the Painlev\'e test can be applied with success. Hodograph
 transformations can be sometimes used to this purpose \cite{CFA89}. In this sense we must say that precisely this
 ubiquity of a PDE, that can appear in many different forms depending of the variables that we have
 choosed, can be solved by means of the SMM. Actually, when we have been able to write a PDE in a form in
 which the SMM has been successful, this method provides us the \textbf{singular manifold equations} that can
 be considered as the \textbf{canonical form} of a PDE. We can conjecture that, if two PDE's have the same singular
 manifold equations, then there exists a transformation that relates the two equations since they are essentially the
 same \cite{e04}.
 \item The usual SMM could be too restrictive when is applied to PDEs with several Painlev\'e branches.
 Modifications of SMM that includes the different branches simultaneously can be found in the following
 references: \cite{estevez93},  \cite{estevez01},
 \cite{e04} and \cite{estevez94}. Once more the solution of this problem includes a bonus: If an equation has two Painlev\'e
 branches, the modification of the SMM provides us not only the right answer but the Miura transformations that
 relate our initial PDE to another PDE with just one Painlev\'e branch (\cite{estevez01}, \cite{e04}).
 \item The SMM requires the truncation of the Painlev\'e series at the constant level $j=a$ and the annulation of
 all the coefficients in the different powers of $\phi$. Sometimes this
 condition is very restrictive and needs to be modified \cite{estevez95}. Specially for some equations in $1+1$ dimensions the SMM
 imposes so many restrictions that there is not freedom enough to get nontrivial solutions and/or to
 introduce a spectral parameter \cite{estevez93}.

 \end{itemize}

 This paper concerns specially with the last of the problems listed above.
  We recall the ARS (Ablowitz, Ramani, Segur) conjecture, \cite{ars}, according to which a PDE has the  Painlev\'e property if all
of its reductions have such a property. Our main aim in this paper
is just the opposite. We show that a PDE in $2+1$ can be much
easier analyzed through the SMM than its reductions to $1+1$. In
fact we present an equation in $2+1$ in which the SMM works very
well. Nevertheless the SMM, when applied directly to its three
simplest reductions to $1+1$ dimensions, presents one or several
of the problems listed above. One can conclude that it is
necessary to increase the number of dimensions in order to have
sufficient freedom for the SMM not be too restrictive.

The plan of the paper is the following: \begin{itemize}

\item In section 2 an equation in $2+1$ dimensions is proposed and
it is proved that  passes the Painlev\'e test. \item In section 3
a complete analysis of the $2+1$ equation is made by means of the
SMM that allows us to obtain the Lax pair and Darboux
transformations for the equation. \item Different reductions are
presented in section 4. Their Lax pairs are also derived by
reduction of the $2+1$ dimensional Lax pair. \item Conclusions are
presented in section 5.
\end{itemize}
\section{An equation in $2+1$ dimensions}
The equation under scrutiny is the following one for a field $h$ depending on  $2+1$ variables $x$, $y$ and
$z$:
\begin{equation}
\left[h_{xxz}-\frac{\textsf{3}}{\textsf{4}}\left(\frac{h_{xz}^2}{h_z}\right)+\textsf{3}h_xh_z\right]_x=h_{yz}.\label{2.1}
\end{equation}
Alternatively we can introduce a new dependent field $p(x,y,z)$ in
order to write (\ref{2.1}) as the following system:
\begin{gather}
  h_z+p^2=0 ,\label{2.2}\\
-p_y+p_{xxx}+\frac{\textsf{3}}{\textsf{2}}\,p\,h_{xx}+\textsf{3}\, p_xh_x=0.\label{2.3}
\end{gather}
We obtained (\ref{2.1}) by searching $2+1$ integrable generalizations of peakon equations. Actually,
(\ref{2.1}) generalizes the Ermakov-Pinney equation \cite{hone99} that, as has been proved in
\cite{honeWang03}, is related by means of a reciprocal transformation to an equation with peakon solutions:
the Degasperis-Procesi equation. Furthermore, (\ref{2.1}) can be considered as a modified version of the
generalized Hirota-Satsuma equation presented in \cite{blmp87} and \cite{estevez95} as a model for an
incompressible fluid.
\subsection{Painlev\'e test}
To check if (\ref{2.1}) passes the Painlev\'e test, the solutions
should be written locally as \cite{WTC}:
\begin{equation}
h=
\sum_{j=0}^{\infty}h_j(x,y,z)\left[\phi(x,y,z)\right]^{j-a}.\label{2.4}
\end{equation}
Substitution of (\ref{2.4}) into (\ref{2.1}) gives us a polynomial
in $\phi$ (we have used MAPLE to handle the calculation)  of the
type:
\begin{equation}
 \sum_{j=0}^{\infty}C_j\left[\phi(x,y,z)\right]^{3j-3a-6}=0\label{2.5}
\end{equation}
The leading term ($j=0$) gives us:
\begin{gather}
a=1,\label{2.6}\\
h_0=\phi_x.\label{2.7}
\end{gather}
The coefficient in $h_j$ is:
\begin{equation}
\phi_x^5\phi_z^3\phi^{j-9}(j-1)(j-3)(j-4)(j+1)\label{2.8}
\end{equation}
which means that the equation has resonances in $j=1,3,4$. It is
not difficult to check that $C_1$, $C_3$ and $C_4$ are identically
$0$ for any value of  $h_1$, $h_3$ and $h_4$. Consequently $h$
admits a   local expansion (\ref{2.4}) in terms of four arbitrary
functions $\phi$, $h_1$, $h_3$ and $h_4$ which means that
\textbf{(\ref{2.1}) passes the Painlev\'e test} \cite{WTC}.
\subsection{Reductions}
There are three obvious reductions of (\ref{2.1}). \begin{itemize}\item 1) $\frac{\partial h}{\partial y}=0$

 Actually it
is equivalent to the reduction $\frac{\partial \hat h}{\partial y}=\frac{\partial \hat h}{\partial x}$ if we
redefine $h$ as $ h=\hat h+\frac{x}{3}$. With this reduction the equation (\ref{2.1}) is
\begin{equation} \left[h_{xxz}-\frac{\textsf{3}}{\textsf{4}}\left(\frac{h_{xz}^2}{h_z}\right)+\textsf{3}h_xh_z\right]_x=0.\label{2.9}\end{equation}
or
\begin{gather}
h_z=-p^2,\notag \\
0=\left(2pp_{xx}-p_x^2+3p^2h_x\right)_x.\label{2.10}
\end{gather}
Integration of (\ref{2.10}) gives us: \begin{equation}\nonumber
pp_{xx}-\frac{p_x^2}{2}+2Vp^2+F(z)=0,\quad\quad V_z=-\frac{3}{4}(p^2)_x\end{equation} that is the
Ermakov-Pinney equation \cite{hone99}. As it has been proved in \cite{honeWang03} (see equation (2.15) of
this reference), this equation arises through a reciprocal transformation from the Degasperis-Procesi
equation \cite{Degproc99} that is an equation with peakon solutions \cite{gp95} similar to the celebrated
Camassa-Holm equation \cite{ch93}. As we see in the next section, the SMM is not effective when applied
directly to (\ref{2.9}). The number of conditions that the SMM requires is so large that only a few trivial
solutions can be identified by this procedure.

\item 2) $\frac{\partial h}{\partial z}=\frac{\partial h}{\partial x}$
This reduction yields trivially to the modified Korteveg de Vries  equation:
\begin{equation} p_y-p_{xxx}+6p^2p_x=0.\label{2.11}\end{equation}
The problem, in this case, is that (\ref{2.11}) has two Painlev\'e branches (see \cite{estevez94}).
Nevertheless, as it has been proved in  \cite{estevez93}, the SMM can be implemented by including both
branches simultaneously. This generalization of the SMM has been applied successfully to many equations with
two Painlev\'e branches \cite{estevez01}, \cite{e04}.
\item 3) $\frac{\partial h}{\partial z}=\frac{\partial h}{\partial y}$
This reduction yields  the $1+1$ equation:
\begin{equation} \left[h_{xxz}-\frac{\textsf{3}}{\textsf{4}}\left(\frac{h_{xz}^2}{h_z}\right)+\textsf{3}h_xh_z\right]_x=h_{zz}.\label{2.12}\end{equation}
or
\begin{gather}
h_z=-p^2,\notag \\
-p_z+p_{xxx}+\frac{\textsf{3}}{\textsf{2}}ph_{xx}+\textsf{3}p_xh_x=0.\label{2.13}
\end{gather}
The problem with the SMM for this equation is exactly the same
that in the case of the reduction 1). The method happens to be too
restrictive and it does not provides a spectral parameter.
\end{itemize}

We see in the next sections how to solve the problems by going to
$2+1$ dimensions.
\section{Singular Manifold Method}
In order to make the calculations much easier, it is convenient to write (\ref{2.1}) in nonlocal form as the
system
\begin{gather}
h_y=n_x,\label{3.1} \\
h_{xxz}h_z-\frac{\textsf{3}}{\textsf{4}}h_{xz}^2+\textsf{3}h_xh_z^2-h_zn_z=0.\label{3.2}
\end{gather}
\subsection{Truncated expansion}
The SSM \cite{Weiss} requires the truncation of the Painlev\'e series  (\ref{2.4}) at the constant level
$j=a$. It implies according to (\ref{2.6})-(\ref{2.7}) that solutions $h^{(1)},n^{(1)}$ \,of (\ref{3.2}) can
be written as:
\begin{gather}
h^{(1)}=  h +\frac{ \phi_x}{ \phi},\notag \\
n^{(1)}=  n +\frac{ \phi_y}{ \phi}.\label{3.3}
\end{gather}
The SMM implies that $h $ and $  n $ are also seminal solutions of
the system (\ref{3.1})-(\ref{3.2}). $ \phi $ is now the so-called
\textbf{singular manifold} associated to the $(h ,n) $ solution.

Substitution of (\ref{3.3}) in (\ref{3.2}) yields a polynomial in
negative powers of $ \phi $ of the form:
\begin{equation}  \sum_{k=0}^4 E_k \left(\frac{\phi_x}{\phi}\right )^{k}=0. \label{3.4}\end{equation}
Setting to zero all the coefficients $E_k$, we get  the following
results (see Appendix A):
\subsection{Seminal solutions}
The seminal field $  h$ can be written in terms of the singular manifold through the following expressions:
\begin{gather}
  h_x=-\frac{  V_x}{3}-\frac{  V^2}{12}+\frac{  Q}{3},\label{3.5} \\
  h_z=-\frac{1}{4  R}\left(  R_x+  R  V\right)^2,\label{3.6}
\end{gather}
where it has been useful to define $  V$, $  R$ and $  Q$ as:
\begin{gather}
  V=\frac{   \phi _{xx}}{  \phi _x},\notag \\
  R=\frac{   \phi _{z}}{  \phi _x},\label{3.7}\\
  Q=\frac{   \phi _{y}}{  \phi _x}.\notag
\end{gather}
The compatibility between the above definitions implies:
\begin{gather}
  V_z=\left(  R_x+  R  V\right)_x\label{3.8}\\
  V_y=\left(  Q_x+  Q  V\right)_x.\label{3.9}
\end{gather}
 It is also convenient to define the Schwartzian
derivative:
\begin{equation}    S=  V_x-\frac{  V^2}{2} \label{3.10}\end{equation}

\subsection{Singular Manifold Equations}
$E_k$=0 provides  (see Appendix A) the following  relation between $  S$, $  Q$ and $  R$.
\begin{equation}    Q_z=  S_z-\frac{3}{2}  R_x\left(  S+\frac{  R_{xx}}{  R}-\frac{  R_{x}^2}{2  R^2}\right) \label{3.11}\end{equation}
that together with (\ref{3.8}) and (\ref{3.9})  constitute the \textbf{singular manifold equations}, which
means the equations that the singular manifold $\phi $ should satisfy in order to have a truncation of the
Painlev\'e series.
\subsection{Lax pair}
The expressions (\ref{3.5})-(\ref{3.6}) can be easily linearized by introducing a new function $  \psi $
defined as:
\begin{equation}    \phi _x=  \psi ^2 \label{3.12}\end{equation}
and consequently according to(\ref{3.7})-(\ref{3.9}), we have:
\begin{gather}
  V=2\frac{  \psi _{x}}{  \psi },\label{3.13} \\
  R_x+  R  V=2\frac{  \psi _{z}}{  \psi },\label{3.14}\\
  Q_x+  Q  V=2\frac{  \psi _{y}}{  \psi },\label{3.15}
\end{gather}

By substituting (\ref{3.13}) and (\ref{3.14})
 into (\ref{3.5})-(\ref{3.6}), we get:
 \begin{gather}
  Q=3  h_x+2\frac{  \psi _{xx}}{  \psi }-\frac{  \psi _x^2}{  \psi ^2},\label{3.16} \\
  R=-\frac{  \psi _z^2}{  h_z \psi ^2}.\label{3.17}
\end{gather}
Substitution of (\ref{3.13}), (\ref{3.16}) and (\ref{3.17}) into
(\ref{3.14})-(\ref{3.15}) provides us the Lax pair:
 \begin{gather}
-  \psi _y+  \psi _{xxx}+3  h_x  \psi _x+\frac{3}{2}  h_{xx}  \psi =0,\label{3.18} \\
  \psi _{xz}-\frac{  h_{xz}}{ 2 h_z}  \psi _z+  h_z  \psi =0,\label{3.19}
\end{gather}
 where the eigenfunction $\psi$ is related to the singular manifold through (\ref{3.12}).

 If we consider the Lax pair as the compatibility condition between the operators
  \begin{gather}
T_1=\partial_x\partial_z-\frac{  h_{xz}}{ 2 h_z} \partial_z+h_z\notag \\
T_2=\partial_y-\partial_x^3-3  h_x \partial_x-\frac{3}{2}  h_{xx}\notag
\end{gather}
and  compare with the spectral problem studied in \cite {{blp87}}
(see equations (2.1) and (2.2) of this reference), it is easy to
see that $T_1$ belongs to the class discussed there, but $T_2$
does not because in our case it has $\partial_x^3$ instead of
$\partial_x^2$ \cite{konopel92}.
\subsection{Darboux Transformations}
Let $\psi_1$ and $\psi_2$ be two different eigenfunctions  for
$h$, which means that they satisfy (\ref{3.18})-(\ref{3.19}).
 \begin{gather}
-  \psi _{1,y}+  \psi _{1,xxx}+3  h_x  \psi _{1,x}+\frac{3}{2}  h_{xx}  \psi_1 =0,\notag \\
  \psi _{1,xz}-\frac{  h_{xz}}{  h_z}  \psi _{1,z}+  h_z  \psi_1 =0.\label{3.20}
\end{gather}
 \begin{gather}
-  \psi _{2,y}+  \psi _{2,xxx}+3  h_x  \psi _{2,x}+\frac{3}{2}  h_{xx}  \psi_2 =0,\notag \\
  \psi _{2,xz}-\frac{  h_{xz}}{  h_z}  \psi _{2,z}+  h_z  \psi_2 =0.\label{3.21}
\end{gather}
Therefore there must be two singular manifolds for $h$ defined as:
\begin{equation}
\phi _{1,x}=\psi _{1}^2,\quad\quad \phi _{2,x}=\psi _{2}^2.\label{3.22}
\end{equation}
According to (\ref{3.3}) we can obtain a new solution $(h^{(1)},\,
n^{(1)})$ through the truncated expansion
\begin{gather}
h^{(1)}=  h +\frac{ \phi_{1,x}}{ \phi_1},\notag \\
n^{(1)}=  n +\frac{ \phi_{1,y}}{ \phi_1}.\label{3.23}
\end{gather}
Since $(h^{(1)},\, n^{(1)})$ is also solution of
(\ref{3.1})-(\ref{3.2}), its Lax pair is:
 \begin{gather}
-  \psi^{(1)} _{y}+  \psi^{(1)} _{xxx}+3  h^{(1)}_x  \psi^{(1)} _{x}+\frac{3}{2}  h^{(1)}_{xx}  \psi^{(1)} =0,\notag \\
  \psi^{(1)} _{xz}-\frac{  h^{(1)}_{xz}}{  h^{(1)}_z}  \psi^{(1)} _{z}+  h^{(1)}_z  \psi^{(1)} =0,\label{3.24}
\end{gather}
where $\psi^{(1)}$ is an eigenfuntion for $h^{(1)}$. So we can
define a singular manifold $\phi^{(1)}$ for $h^{(1)}$ through the
following expression
\begin{equation}\phi^{(1)}_x=\left(\psi^{(1)}\right)^2.\label{3.25}\end{equation}
The idea is to consider (\ref{3.1}), (\ref{3.2}), (\ref{3.24}) and (\ref{3.25}) as a nonlinear system of PDEs
in $h^{(1)}$, $n^{(1)}$, $\psi^{(1)}$ and $\phi^{(1)}$. Therefore the truncated expansion (\ref{3.23}) for
$h^{(1)}$ and  $n^{(1)}$ can be extended to $\psi^{(1)}$ and $\phi^{(1)}$ as:
\begin{gather}
\psi^{(1)}=  \psi_2 +\frac{\Lambda}{ \phi_1},\notag \\
\phi^{(1)}=  \phi_2 +\frac{\Delta}{ \phi_1}.\label{3.26}
\end{gather}
Substitution of (\ref{3.23}) and (\ref{3.26}) in (\ref{3.24}) and (\ref{3.25}) provides us polynomials in
negative powers of $\phi_1$. The result (see Appendix B) is:
\begin{gather}
\Lambda= -\psi_1\Omega,\notag \\
\Delta= -\Omega^2,\label{3.27}
\end{gather}
where $\Omega$ satisfies : \begin{equation}d\Omega=
\psi_1\psi_2\,dx+\left(\psi_1\psi_{2,xx}+\psi_2\psi_{1,xx}-\psi_{1,x}\psi_{2,x}+3h_x\psi_1\psi_2\right)\,dy-\frac{\psi_{1,z}\psi_{2,z}}{h_z}\,dz\label{3.28}\end{equation}
Therefore (\ref{3.23}) and (\ref{3.26}) are binary Darboux transformations in the sense that they allow us to
construct the iterated Lax pair (\ref{3.24}) through the solutions of two seminal Lax pairs (\ref{3.20}) and
(\ref{3.21}). Nevertheless we should remark that these transformations are not the usual binary Darboux
transformations that appear, for instance, in references \cite{lr84} and \cite{ms91}, because in (\ref{3.26})
not only the eigenfunctions $\psi_1$ and $\psi_2$ appear but also the singular manifold $\phi_1=\int
\psi_1^2\,dx$. Transformations like (\ref{3.23})-(\ref{3.26}) have been denominated B\"acklund-gauge
transformations in reference \cite{ks91}.
\subsection{Iterated solutions}
As we have shown above, $\phi^{(1)}$ is a singular manifold for
$h^{(1)}$. Therefore it can be used to iterate (\ref{3.23}) in the
following form:
\begin{gather}
h^{(2)}=  h^{(1)} +\frac{ \phi^{(1)}_{x}}{ \phi^{(1)}},\notag \\
n^{(2)}=  n^{(1)} +\frac{ \phi^{(1)}_{y}}{ \phi^{(1)}}.\label{3.29}
\end{gather}
This second iteration (\ref{3.29}) can be combined with the first
one (\ref{3.23}),  and with  (\ref{3.26}) and  (\ref{3.27}) to
give:
\begin{gather}
h^{(2)}=  h +\frac{ \tau_{x}}{ \tau},\notag \\
n^{(2)}=  n +\frac{ \tau_{y}}{ \tau},\label{3.30}
\end{gather}
where
\begin{equation}\tau=\phi^{(1)}\phi_1=\phi_2\phi_1-\Omega^2.\label{3.31}\end{equation}
\section{Reductions}
As we said in subsection 2.2, there are several interesting reductions of (\ref{2.1}).
\subsection{$\frac{\partial h}{\partial y}=0$}
The equation can be written as:
\begin{equation}
\left[h_{xxz}-\frac{\textsf{3}}{\textsf{4}}\left(\frac{h_{xz}^2}{h_z}\right)+\textsf{3}h_xh_z\right]_x=0\label{4.1}
\end{equation}
\begin{gather}
or:
h_z=-p^2,\notag \\
\left(2pp_{xx}-p_x^2+3p^2h_x\right)_x=0,\label{4.2}
\end{gather}
that, after integration with respect to $x$,  is the Ermakov-Pinney equation $\quad
pp_{xx}-\frac{p_x^2}{2}+2Vp^2+F(z)=0$ with $V_z=-\frac{3}{4} (p^2)_x$ (see  \cite{hone99} and
\cite{honeWang03}). This reduction implies $Q=0$. To get a right Lax pair with a spectral parameter it is
necessary to go to the $2+1$ Lax pair (\ref{3.18})-(\ref{3.19}) and make the gauge-reduction
$\psi(x,y,z)=e^{\lambda y}\hat\psi(x,z)$. The reduction of (\ref{3.18})-(\ref{3.19}) is obviously:
 \begin{gather}
  \hat\psi _{xxx}+3  h_x  \hat\psi _x+\left(\frac{3}{2}  h_{xx}-\lambda\right)  \hat\psi =0,\label{4.3} \\
  \hat\psi _{xz}-\frac{  h_{xz}}{  h_z}  \hat\psi _x+  h_z \hat \psi =0.\label{4.4}
\end{gather}
The compatibility condition between (\ref{4.3})and (\ref{4.4}) gives us the \textbf{third-order spectral
problem}
 \begin{gather}
 \hat \psi _{xxx}+3  h_x  \hat\psi _x+\left(\frac{3}{2}  h_{xx}-\lambda\right)  \hat\psi =0, \notag\\
\lambda\hat\psi_z=-h_z\psi_{xx}+\frac{h_{xz}}{2}\hat\psi_x+\left(-\frac{h_{xxz}}{2}+\frac{h_{xz}^2}{4h_z}-3h_xh_z\right)\hat\psi=0,\label{4.5}
\end{gather}
that is the Lax pair of reference \cite{hone02}.
\subsection{$\frac{\partial h}{\partial z}=\frac{\partial h}{\partial x}$}
In this case the reduction of (\ref{2.2}) is:
\begin{equation} h_x=-p^2.\label{4.6}\end{equation}
Therefore (\ref{2.3}) is the modified Korteveg-de Vries equation:
 \begin{equation}-p_y+p_{xxx}-6p^2p_x=0.\label{4.7}\end{equation}
Before doing the reduction in the Lax pair, we need to introduce the spectral parameter through the gauge
$\psi=e^{\lambda z}\hat\psi$ and then make the reduction $\frac{\partial h}{\partial z}=\frac{\partial h
}{\partial x}$, $h_x=-p^2$ in (\ref{3.18})-(\ref{3.19}). The result is the \textbf{second-order spectral
problem}:
 \begin{gather}
  \hat\psi _{xx}+\left(\lambda-\frac{p_x}{p}\right)\hat\psi _x+\left(-p^2-\lambda\frac{p_x}{p}\right)
   \hat\psi =0,\notag\\
  \hat\psi _{y}+\left(2p^2+\lambda\frac{p_x}{p}-\frac{p_{xx}}{p}-\lambda^2\right)\hat\psi _x+\left(\-\lambda\frac{p_{xx}}{p}+
  \lambda^2\frac{p_x}{p}+\lambda p^2\right)  \hat\psi =0.\label{4.8}
\end{gather}

\subsection{$\frac{\partial h}{\partial z}=\frac{\partial h}{\partial y}$}
The reduction gives us the $1+1$ PDE
\begin{equation} \left[h_{xxz}-\frac{\textsf{3}}{\textsf{4}}\left(\frac{h_{xz}^2}{h_z}\right)+\textsf{3}h_xh_z\right]_x=h_{zz}.\label{4.9}
\end{equation}
To have the right reduction of the Lax pair, we need to make the gauge transformation $\psi=e^{\lambda
y}\hat\psi$ and the reduction $\frac{\partial h}{\partial z}=\frac{\partial h}{\partial y}$. In this case
(\ref{3.18})-(\ref{3.19}) reduce to:
 \begin{gather}
  -\hat \psi_z+\hat\psi _{xxx}+3  h_x  \hat\psi _x+\left(\frac{3}{2}  h_{xx}-\lambda\right)  \hat\psi =0,\label{4.10} \\
  \hat\psi _{xz}-\frac{  h_{xz}}{  h_z}  \hat\psi _x+  h_z \hat \psi =0.\label{4.11}
\end{gather}
Solving (\ref{4.10}) for $\hat\psi_z$ and substituting this into
(\ref{4.11}), we get the \textbf{fourth-order spectral problem}:
 \begin{gather}
 \hat\psi _{xxxx}-\frac{h_{xz}}{2h_z} \hat\psi _{xxx}+3h_x \hat\psi _{xx}+\notag\\\quad \quad\quad+\left( \frac{9h_{xx}}{2}-\frac{3h_xh_{xz}}{2h_z}
-
 \lambda\right)\hat\psi _{x}+\left(h_z+\frac{3h_{xxx}}{2}
 +\frac{\lambda h_{xz}}{2h_z} -\frac{3h_{xz}h_{xx}}{4h_z}\right)\hat\psi=0,\notag \\
  -\hat \psi_z+\hat\psi _{xxx}+3  h_x  \hat\psi _x+\left(\frac{3}{2}  h_{xx}-\lambda\right)  \hat\psi =0.\label{4.12}
\end{gather}

\section{Conclusions}
\begin{itemize}
\item A new equation (2.1) in $2+1$ dimensions is presented and
the Painlev\'e test is successfully applied. \item  It should be
noted that, although the SMM presents several problems when it is
applied directly to different $1+1$ dimensional reductions of
(2.1), it  has been proved to be an excellent  tool to analyze the
$2+1$ dimensional equation (2.1). It works very well and allow us
to obtain  the Lax pair, Darboux transformations and an iterative
method to obtain solutions of (2.1). \item The Lax pairs for the
different $1+1$ dimensional reductions can be derived from the
$2+1$ dimensional Lax pair.

\end{itemize}

\section*{Acknowledgements}
This research has been supported in part by DGICYT under projects BFM2002-02609 and BFM2000-1327.

\section*{Appendix A}
 We used MAPLE to compute the polynomial that results from the substitution of
(\ref{3.3}) into (\ref{3.2}): The result is a polynomial of the
form (\ref{3.4}) the first coefficient of which is:

\begin{equation}E_4=\frac{R^2}{4}\left(6h_x+3v_x-S-2Q\right)\label{5.1}\end{equation}
that can be solved as:
\begin{equation}h_x=-\frac{V_x}{2}+\frac{Q}{3}+\frac{S}{6}.\label{5.2}\end{equation}
At this point it is useful to remember that the Painlev\'e
Property is invariant under homographic transformations
\cite{Weiss}. Therefore it is convenient to write everything in
terms of the homographic invariants $S$, $R$ and $Q$. It can be
done if we introduce the change:
 \begin{gather}
h=\alpha-\frac{V}{2},\notag \\
 n=\beta-\frac{Q_x+QV}{2}.\label{5.3}
\end{gather}
From (\ref{5.2}) and (\ref{5.3}) we have that $\alpha_x$ is the
homographic invariant
\begin{equation}\alpha_x=\frac{Q}{3}+\frac{S}{6}\label{5.4}\end{equation}
Substitution of (\ref{5.2}) and (\ref{5.3}) into (\ref{3.4}) gives
us:
\begin{equation}E_3=0\notag\end{equation}
\begin{equation}E_2=
\frac{1}{6}\left(-18\alpha_z^2+6R\alpha_z\left(S-Q\right)+6R\beta_z+2RS_{xz}-5RQ_{xz}\right)\label{5.5}\end{equation}
\begin{equation}E_1=
-E_2V-\frac{1}{6}\left(6\alpha_z(Q_z-S_z+SR_x-QR_x)-R_x(2S_{xz}-5Q_{xz})+6R_x\beta_z\right).\label{5.6}\end{equation}
We can solve (\ref{5.5}) for $\beta_z$ and by substituting it into
(\ref{5.6}) we have:
\begin{equation}\alpha_z=
\frac{R}{3R_x}\left(S_z-Q_z\right).\label{5.7}\end{equation} The
compatibility condition $\alpha_{xz}=\alpha_{zx}$ between
(\ref{5.4}) and (\ref{5.7}) provides us:
\begin{equation}\left(S_z-Q_z\right)^2\left(-4R^2\left(S_z-Q_z\right)+R_x\left(-3R_x^2+6RR_{xx}+6SR^2\right)\right)=0.\label{5.8}\end{equation}
The solution $S_z=Q_z$ of (\ref{5.8}) it is not useful because in such a case (\ref{5.7}) implies
$\alpha_z=0$. Therefore the solution of (\ref{5.8}) should be:
\begin{equation}    Q_z=  S_z+\frac{3}{2}  R_x\left(  S+\frac{  R_{xx}}{  R}-\frac{  R_{x}^2}{2  R^2}\right),\label{5.9}\end{equation}
that is the singular manifold equation (\ref{3.11}). Substitution
of (\ref{5.9}) into (\ref{5.7}) yields:
\begin{equation}\alpha_z=
\frac{R_{xx}}{2}+\frac{RS}{2}-\frac{R_{x}^2}{4R}.\label{5.10}\end{equation}
Substitution of (\ref{5.4}) and (\ref{5.10}) into (\ref{5.3})
gives finally us (\ref{3.5})-(\ref{3.6}).
\section*{Appendix B}
$\bullet$ Substitution of  (\ref{3.26}) into the second of equations (\ref{3.24}) gives us the following
polynomial in $\phi_1$
\begin{equation}\left(\phi_{2,x}-\psi_2^2\right)+\frac{1}{\phi_1}\left(\Delta_x-2\psi_2\Lambda\right)-
\frac{1}{\phi_1^2}\left(\Delta\phi_{1,x}+\Lambda^2\right)=0.\label{6.1}\end{equation} By using (\ref{3.22})
we get:
\begin{equation}\frac{1}{\phi_1}\left(\Delta_x-2\psi_2\Lambda\right)-\frac{1}{\phi_1^2}\left(\Delta\psi_{1}^2+\Lambda^2\right)=0.\label{6.2}\end{equation}
Setting to zero both coefficients we have:
 \begin{gather}
\Delta =-\left(\frac{\Lambda}{\psi_1}\right)^2,\notag \\
\left(\frac{\Lambda}{\psi_1}\right)_x+\psi_1\psi_2=0,\label{6.3}
\end{gather} that can be easily written as:
 \begin{gather}
\Delta =-\Omega^2,\notag \\
\Omega_x=\psi_1\psi_2,\label{6.4}
\end{gather}
by introducing
\begin{equation}\Omega=-\frac{\Delta}{\psi_1}.\label{6.5}\end{equation}
$\bullet$ Substitution of (\ref{3.26}) into the second of the equations (\ref{3.24}) gives us:
\begin{equation}\left(\psi_{1,x}-\psi_1\frac{h_{xz}}{2h_z}\right)
\left[\frac{1}{\phi_1}\left(\Omega_zh_z+\psi_{1,z}\psi_{2,z}\right)-\frac{1}{\phi_1^2}\left(\Omega
h_z\phi_{1,z}+\Omega\psi_{1,z}^2+\phi_{1,z}\psi_{2,z}\psi_1-\Omega_z\psi_{1,z}\psi_1\right)\right],\label{6.6}\end{equation}
where we have used (\ref{6.4}) and $\psi_{i,xz}=-h_z\psi_i+{\displaystyle\frac{h_{xz}}{h_z}\psi_{i,z}},\quad
i=1,2.$

From (\ref{3.6}), (\ref{3.7}) and  (\ref{3.17})  we have:
\begin{equation}\phi_{1,z}= -\frac{\psi_{1,z}^2}{h_z}\label{6.7}\end{equation}
that can be substituted into  (\ref{6.6}) to give:
\begin{equation}\left(\psi_{1,x}-\psi_1\frac{h_{xz}}{2h_z}\right)\left(\frac{1}{\phi_1}+\frac{\psi_1\psi_{1,z}}{h_z\phi_1^2
}\right)\left(\Omega_zh_z+\psi_{1,z}\psi_{2,z}\right). \label{6.8}\end{equation} Therefore
\begin{equation}\Omega_z=-\frac{\psi_{1,z}\psi_{2,z}}{h_z} \label{6.9}\end{equation}
$\bullet$ Substitution of  (\ref{3.26}) in the first of the equations (\ref{3.25}) gives us the following
polynomial in $\phi_1$
\begin{gather}\frac{\psi_1\Omega}{\phi_1^2}\left(2\psi_1\psi_{1,xx}-
\psi_{1,x}^2+3h_x\psi_1^2-\phi_{1,y}\right)-\frac{\psi_1}{\phi_1}\Omega\left(\psi_{1,y}-\psi_{1,xxx}-3h_x\psi_{1,x}-\frac{3}{2}h_{xx}\psi_1\right)\notag\\-\frac{\psi_1}{\phi_1}\left(\psi_1\psi_{2,xx}+\psi_2\psi_{1,xx}-
\psi_{1,x}\psi_{2,x}+3h_x\psi_1-\Omega_y\right)=0.\label{6.10}\end{gather} From (\ref{3.5}) and (\ref{3.7}) we
have:
\begin{equation}\phi_{1,y}=\phi_{1,x}\left(3h_x+V_{1,x}+\frac{V_1^2}{4}\right)\label{6.11}\end{equation}
and, if we use (\ref{3.12}) and (\ref{3.13}),
\begin{equation}\phi_{1,y}=3h_x\psi_1^2+2\psi_1\psi_{1,xx}-\psi_{1,x}^2.\label{6.12}\end{equation}
Substitution of (\ref{6.12}) and (\ref{3.20}) into (\ref{6.10})  provides us:
\begin{gather}-\frac{\psi_1}{\phi_1}\left(\psi_1\psi_{2,xx}+\psi_2\psi_{1,xx}-
\psi_{1,x}\psi_{2,x}+3h_x\psi_1-\Omega_y\right)=0\label{6.13}\end{gather} and therefore:
\begin{equation}\Omega_y=\psi_1\psi_{2,xx}+\psi_2\psi_{1,xx}-
\psi_{1,x}\psi_{2,x}+3h_x\psi_1.\label{6.14}\end{equation}

\label{lastpage}

\end{document}